\documentclass[12pt]{article}
\textwidth
16.4cm
\oddsidemargin
2.5cm
\advance\oddsidemargin by
-1in
\evensidemargin
0.0cm
\advance\evensidemargin
by
-1in
\marginparwidth
1.9cm
\marginparsep
0.4cm
\marginparpush
0.4cm
\topmargin
-1.5cm
\advance\topmargin
by
-0.0in
\textheight
22.5cm
\makeindex

\newcommand\noi{\noindent}
\newcommand\beq{\begin{equation}}
\newcommand\eeq{\end{equation}}
\newcommand\beqn{\begin{eqnarray}}
\newcommand\eeqn{\end{eqnarray}}
\newcommand{\la}{\langle}
\newcommand{\ra}{\rangle}
\newcommand{\doublespace}
{
\renewcommand{\baselinestretch}
{1.6}
\large\normalsize}

\begin{document}


\doublespace

\vspace*{3cm}
\centerline{\Large \bf Soft QCD Dynamics of Elastic Scattering}

\medskip

\centerline{\Large \bf in Impact
Parameter Representation}

\vspace{.5cm}
\begin{center}
 {\large B.Z.~Kopeliovich$^{1,2}$, I.K. Potashnikova$^{1,2}$, 
B.~Povh$^{1}$
and E.~Predazzi$^{3}$}

\vspace{0.3cm}

$^{1}$ {\sl Max-Planck Institut
f\"ur
Kernphysik,
Postfach
103980, 69029 Heidelberg,
Germany}\\

$^{2}${\sl Joint
Institute
for Nuclear Research, Dubna,
141980
Moscow Region,
Russia}\\

$^3${\sl Universit\`a di
Torino
and INFN, Sezione di Torino, I-10125, Torino,
Italy}

\end{center}

 \vspace{1cm}
 \begin{abstract} 
 The elastic hadronic amplitude is calculated using the nonperturbative
light-cone dipole representation for gluon bremsstrahlung. The data for large
mass diffraction demand a two-scale structure of light hadrons: the gluon
clouds of the valence quarks with the size of $\sim 0.3\,fm$ and the hadronic
size $\sim 1\,fm$. The presence of the two scales unavoidably leads to a
specific form for the total hadronic cross section which consists of a steeply
rising $\propto s^{\Delta}$ ($\Delta=0.17\pm 0.01$) term related to gluon
radiation, and a large constant term originating from soft interactions which
does not induce any gluon emission. Our calculations reproduce well the total
cross sections and elastic slopes \cite{k3p}. To further test the model, we
analyze the elastic $pp$ and $\bar pp$ differential cross sections and extract
the partial amplitudes in the impact parameter representation. The Pomeron
trajectory as a function of the impact parameter is only slightly above one for
central collisions, but steeply grows towards the periphery. The model predicts
correctly the shape and energy dependence of the partial amplitude at all
impact parameters.

\end{abstract}

\bigskip
\doublespace
\newpage
\noi
\section{Introduction}

\subsection{Rising total cross sections}

The increase of the total hadronic cross section at high energies
is well known since the ISR experiments in the early 70's. This
discovery came just in the time when the Regge theory had its
conjecture. The simple idea to shift the intercept of the
Pomeron pole above one, $\alpha_P(0)=1+\Delta$, leads to
contradiction with the unitarity restrictions, in particular, the
Froissart bound is violated. It took special efforts to formulate
a self-consistent Regge scheme \cite{dklt} in which unitarity is
restored via Regge cuts and without violation of energy conservation
(warnings of which had been given in \cite{agk}).

The assumption that the Pomeron which governs the hadronic
elastic amplitude at high energies is a Regge pole has no
theoretical justification beyond simplicity. It faces problems
interpreting data from HERA which demonstrate that $\Delta$
substantially increases with $Q^2$ (see,
however, \cite{ck}, \cite{bgp}).

The ensemble of data on hadronic elastic scattering at small $t$, {\it
i.e.} total and differential cross sections, slopes and ratios of real to
imaginary part of the forward elastic amplitudes can successfully be
fitted by many phenomenological models based on varieties of quite
different assumptions aimed to fit the data (see, for example
\cite{dgmp}).  Even the simple parameterization $s^{\Delta}$ with
exponential $t$-dependence for the elastic amplitude describes well the
data at small $t$ \cite{dl,pdt}.  However, the imaginary part of the
forward elastic amplitude is connected by the unitarity relation to the
total inelastic cross section, in other words, the Pomeron is the {\it
shadow} of inelastic processes. Unavoidably, one should study the
dynamics of inelastic collisions to understand the forward elastic
scattering \cite{agk}, rather than guessing its analytic form which is
only mildly restricted by general principles.

The total cross section for a highly virtual photon interacting with a proton
measured in deep-inelastic lepton scattering (DIS) can be estimated using
perturbative QCD if the photon virtuality $Q^2\gg Q_0^2$ ($Q_0\sim 1\,GeV$) and
the energy $s\gg s_0$ ($s_0\sim 1\,GeV^2$), but $x=Q^2/s\ll 1$. Depending on
the approximations used, two models for the hard Pomeron are known: the BFKL
\cite{bfkl} and the double-leading-log DGLAP (see in \cite{book,review}). The
rising energy dependence is interpreted perturbatively as caused by gluon
bremsstrahlung with growing phase space for radiated gluons.  The total cross
section is predicted to rise steeply with energy as is confirmed by data from
HERA. The energy dependence, parameterized as $s^{\Delta_{eff}(Q^2)}$ reveals
the exponent to increase with $Q^2$ up to $\Delta_{eff}\sim 0.5$.

In terms of the QCD light-cone dipole approach one can treat DIS at small
$x$ as an interaction of a tiny size, $\sim 1/Q$, quark-antiquark
fluctuations surrounded by a gluon cloud which is much larger
(logarithmically) than the $\bar qq$ pair.

\subsection{Soft interaction limit: the two scales for light hadrons}

\begin{figure}[tbh]
\includegraphics{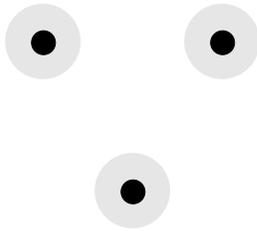} 
\begin{center} 
\vspace{4cm} 
\parbox{13cm} 
{\caption[Delta]
{\sl A skeleton of three valence quarks in the proton, surrounded by
gluon clouds of much smaller size than the mean quark separation.}
\label{3q}} 
\end{center} 
\end{figure}

A new scheme for performing explicit
calculations for the interaction of light hadrons has been suggested in
\cite{k3p}.  It exploits the smallness of the gluon correlation radius
which has been estimated in many approaches. 
In particular, the model developed in ref. \cite{kst2}, extends
the perturbative methods of the light-cone QCD to the nonperturbative
region introducing the light-cone potential into the Schr\"odinger
equation for the Green function that describes the propagation of a
quark-gluon interacting pair. The interaction potential fixed by the data
for large mass soft diffraction turns out to be rather strong leading also
to a short separation $r_0 = 0.3\,fm$ between gluons and the source (a
quark or a gluon).  This result is confirmed by the recent analysis
\cite{hs} of HERA data for diffraction which leads to an even smaller
estimate $r_0\approx 0.2\,fm$ (but with large uncertainties). Thus, a
proton looks in the infinite momentum frame like three valence quarks
surrounded by small gluon clouds as illustrated in Fig.~\ref{3q}.

Such a two-scale structure of light hadrons appears not only in the model
\cite{kst2} treating the vacuum fluctuations as Weizs\"aker-Williams
gluons.  The smallness of the gluon clouds of the valence quarks is
confirmed by the study of the gluon formfactor of the proton employing QCD
sum rules \cite{braun}. The $Q^2$ dependence of the formfactor turns out
to be rather weak corresponding to a small radius of the gluon
distribution which was estimated at the same value $r_0\approx 0.3\,fm$.
The small gluon correlation radius $\sim 0.3\,fm$ appears also from
lattice calculations \cite{pisa}. It is also predicted by the liquid
instanton model \cite{shuryak1,shuryak2} and is related to the instanton
size $\rho_0 \sim 0.3\,fm$. The experimental observation of a small cross
section for large mass soft diffractive dissociation has led to a small
value of $r_0$ \cite{kst2,hs} and this, quite in general, can be taken as
the confirmation of the small size cloud of any kind of gluonic vacuum
fluctuations surrounding the valence quarks. These are usually referred to
as {\it constituent quarks} although nothing specific (about mass,
additivity, etc.) is assumed beyond the simple statement that the clouds
of vacuum fluctuations dressing the valence quarks are much smaller than
the mean hadronic radius.

Of course the transverse size of the gluonic spots increases with energy since
the weights of higher Fock components grow as powers of ${\rm} ln(s)$
\cite{k3p,shuryak2}. Such a behavior is specific of gluonic fluctuations.
Nevertheless, the mean size of the fluctuation clouds is still small compared
to the radii of light hadrons in the energy range of modern accelerators (see
\cite{k3p} and below). The ratio of the constituent quark radius $r_0$ to the
mean interquark separation $R_h$ squared serves as a small number.
Correspondingly, one should single out two different contributions to the total
inelastic cross sections.

The first one is due to the soft interaction which is unable to resolve the
structure of the constituent quarks and excite them. This contribution can be
treated as the cross section of interacting hadrons made of structureless
valence quarks which can be viewed as the {\it skeleton} of the hadrons. One
can try to evaluate it using either the naive two-gluon exchange approximation
\cite{l,n,gs,kl,shuryak2,kkl} or more sophisticated nonperturbative approaches
like string crossing and rearrangement \cite{gn}, or the interaction of
overlapping Wilson loops in the stochastic vacuum model \cite{dosch,pirner}.
This part of the cross section $\sigma_0(R_h)$ is controlled by the mean
interquark separation $R_h$ and is independent of energy since the size of the
quark skeleton of the hadron is constant.

The second contribution to the total cross section comes from the
semi-hard interaction able to resolve the small size of the constituent
quarks and to excite them giving origin to gluon radiation. This cross
section is proportional to the size of the constituent quark, $\propto
r_0^2$. The radiation of each new gluon leads as usual to an extra power
of ${\rm}ln(s)$ which exponentiates to a $\sigma_1(r_0)\,s^{\Delta}$
dependence. However, the energy independent term $\sigma_1(r_0)$ needed
for the exponentiation of these logs is rather small ($\propto r_0^2$) and
can not match the large term $\sigma_0(R_h)$.

Thus, we arrive at the following general structure of the total cross
section which corresponds to a two-scales scheme for the hadronic
structure,
 \beq
\sigma_{tot}\ =\ \tilde\sigma_0\ +\  
\sigma_1(r_0)\,\left(\frac{s}{s_0}\right)^{\Delta}\ ,
\label{stot}
\eeq
 where $\tilde\sigma_0 = \sigma_0(R_h)-\sigma_1(r_0)$. Parametrically,
$\sigma_0(R_h) \gg \sigma_1(r_0)$.

The double scale structure of light hadrons, ({\i.e.} small constituent
quarks versus large interquark separation), leads to the structure of the
total cross section Eq.~(\ref{stot}), rather more complex than the usually
assumed overall behavior $\propto s^{\Delta}$. Of course, in the spirit of
the leading-log(s) approximation, one can neglect the constant term
$\sigma_0$ as $s\to\infty$, but then $\Delta$ should not be compared with
the experimental data available in the energy range where $\sigma_0$ gives
an important contribution. The effective slope of the energy dependence
may, in fact, be substantially smaller,
 \beq \Delta_{eff} = 
\left(1 - \frac{\tilde\sigma_0}
{\sigma_{tot}}\right)\,\Delta\ .
\label{del-eff}
\eeq 
 In fact,  in \cite{k3p} it is found that $\Delta_{eff}\approx
\Delta/2$.

\subsection{Outline of the paper}

This paper is organized as follows. In section~\ref{radiation} we
calculate the cross section of gluon radiation in high energy hadronic
interactions. It is shown that valence quarks contribute additively to the
radiation cross section due to the short-range correlation of the radiated
gluons. Moreover, the sum of the multigluon radiation cross sections which
depend on energy as powers of ${\rm ln}s$ exponentiates in the leading-log
approximation to the energy dependent power $s^{\Delta}$.  The exponent
proportional to the running QCD coupling $\alpha_s$ turns out to be rather
large, $\Delta= 0.17\pm 0.01$ compared to what is believed to be demanded
by the present data for total cross section. However, gluon radiation
contributes with a rather small factor proportional to $r_0^2 \approx
1\,mb$. This energy dependent fraction of the total cross section is fully
predicted. The large energy independent part of the cross section is due
to the interaction of the valence quarks with no gluon radiation. This
cross section is related to the large hadronic size, rather than to $r_0$,
and cannot be evaluated perturbatively. Although it can be estimated in
models, {\it e.g.} as it is done in the stochastic vacuum model
\cite{dosch,pirner}, the uncertainty of such calculations is too large,
and we prefer to treat $\tilde\sigma_0$ as a free parameter, which turns
out to be the only unknown of the model.

The rising total cross section eventually violates the Froissart-Marten
bound at very high energies, but the partial elastic amplitude at small
impact parameters is already very close to the limit imposed by unitarity
and may easily break it down. The procedure of unitarization of the
elastic partial amplitudes is described in section~\ref{unitarization}.
We use the standard quasi-eikonal model, but we compare also with a
different QCD motivated approach.

The model is analyzed with respect to the total cross section data in
section~\ref{forward}. The only parameter of the model, $\tilde\sigma_0$,
can be fixed by comparison with the data at any chosen energy. Then, the
energy dependence is predicted in a good agreement with the data. The
slope of forward elastic scattering needs no new parameters and is also
well predicted.

In the standard Regge phenomenology the energy dependence of the total
cross sections and of the elastic slopes are controlled by the intercept
and slope of the Pomeron trajectory, respectively, which are independent
parameters. However, one may expect them to be correlated since in QCD
the cross sections depend on the hadronic sizes due to color screening
\cite{l,n,gs,hp}. An attempt to incorporate this property was made
recently in \cite{p}. Here we develop this approach treating more
consistently the phase space for the radiated gluons.

The comparison with the data turns out to be most effective in the impact
parameter representation. First of all, the radius of interaction exposes
explicitly in this case. Secondly, unitarity imposes severe restrictions
on the elastic partial amplitude for central hadronic collisions
\cite{kpp}, which slows down the energy dependence of the partial
amplitude \cite{as,kpp}. Thirdly, the color dipole representation in QCD
introduced in \cite{zkl} became a popular tool to study high energy QCD
dynamics in DIS, Drell-Yan reaction etc., since color dipoles are the
eigenstates of the interaction at high energies. In this respect, the
impact parameter representation is suitable for a direct comparison of the
data with a dynamical model (see for instance \cite{soffer}). And, last
but not least, the shape of the amplitude in the impact parameter space is
related to the shape of the amplitude as function of momentum transfer in
a wide range of $t$, rather than only in the forward direction.

In section~\ref{impact} we analyze the available high energy data from
$ISR$ and $S\bar ppS$ for $pp$ and $\bar pp$ elastic scattering to
extract the partial elastic amplitude in the impact parameter
representation.  We follow the procedure suggested by Amaldi and Schubert
\cite{as} who performed a similar analysis of ISR data and concluded that
the total cross section rises due to peripheral interactions while the
partial amplitude for central collisions is energy independent. This is
usually treated as a manifestation of unitarity saturation.  However, the
parameterization of the amplitude used in \cite{as} was based on the
geometrical scaling model which assumes that the ratio of the total cross
section to the elastic slope is independent of energy.  This assumption
unavoidably leads to a constant partial amplitude at zero impact
parameter (see Fig.~\ref{delta}).

The geometrical scaling is known to be broken beyond the ISR energy
range, therefore we perform the analysis differently, in a less model
dependent way. We fit the $t$-dependence of the cross section
independently at each energy assuming no correlation between different
energies, except the normalization which is adjusted to fit the energy
dependent total cross section and ratio of real to imaginary parts of the
forward elastic amplitude. The $t$-dependent imaginary part of the
amplitude arising from the fit is then Fourier transformed to the impact
parameter representation at each energy. The $b$-dependence of the
partial amplitudes found this way is very close to what our model
predicts. Not only the shape of the partial amplitude is well reproduced,
but also its development as function of energy.

In section~\ref{trajectory} we compare the data for the partial amplitudes
at different energies and conclude that they hardly vary at $b\approx 0$,
but rise steeply with energy at large $b > 1\,fm$.  The effective Pomeron
trajectory is a steeply rising function of the impact parameter. Our model
correctly predicts this dependence.

The results of the paper are summarized in section~\ref{summary}.  
Further evidences of the large value of $\Delta$ suggested by data on
diffraction in DIS and particle production at mid rapidities in soft
hadronic collisions are reviewed.

\section{Excitation of valence quarks: nonperturbative gluon
radiation}\label{radiation}

To calculate the energy dependent total cross section one should sum up
the various contributions of different Fock components of the incoming
hadron. To avoid double-counting, we sum the cross sections $\sigma_n$ of
physical process of radiation of $n$ gluons,

\beq
\sigma^{hN}_{tot}=\sum\limits_{n}\sigma^{hN}_n\ .
\label{6.3a}
\eeq

The lowest Fock component ($n=0$) of a hadron 
contains only valence quarks.
For the sake of simplicity we assume the beam hadron to be a meson,
the generalization to a nucleon is simple and is done below.

\begin{figure}[tbh]
\includegraphics{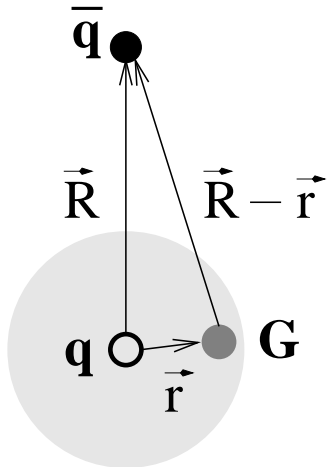}
\begin{center}
\vspace{6cm}
\parbox{13cm}
{\caption[Delta]
{\sl A cartoon for the Fock state $|\bar qqG\ra$
in the impact parameter plane.}
\label{qqG}}
\end{center}
\end{figure}

The contribution to the total cross section corresponding to the
interaction without any gluon radiation has the form,
 \beq
\sigma^{hN}_0 = \int\limits_0^1 d\alpha_q
\int d^2 R\,\left|\Psi^h_{\bar qq}(\alpha_q,R)\right|^2\,
\sigma^N_{\bar qq}(R)\ .
\label{6.4}
\eeq
 Here the valence quark wave function of the hadron, $\Psi^h_{\bar
qq}(\alpha_q,R)$, depends on the transverse $q-\bar q$ separation $R$ (see
Fig.~\ref{qqG})
 and the fraction $\alpha_q$ of the light-cone momentum of the pair carried
by the quark. The energy independent Born cross section of interaction of a
large $\bar qq$ dipole with a nucleon $\sigma^N_{\bar qq}(R)$ cannot be
calculated perturbatively since the separation $R$ is large. It can not be
adjusted directly to the experimental data since the data include the
contribution from gluon bremsstrahlung leading to the energy dependence of
$\sigma^N_{\bar qq}$. Instead, we treat $\sigma^{hN}_0$ as a free
parameter.

The next contribution to the $\sigma_{tot}^{hN}$ comes from
radiating a single gluon. The radiation is possible only due
to the difference between the amplitudes for the $\bar qq$ and $\bar
qqG$ Fock components, otherwise the interaction does not alter the
combination of Fock states and they remain coherent, {\it
i.e.} nothing new is produced. Another way to explain this is to say
that the interaction with the target can free the gluon
fluctuation only if it resolves it, {\it i.e.} discriminates
between the interaction amplitudes for two Fock components $|\bar
qq\ra$ and $|\bar qqG\ra$.

The contribution to the total cross section
corresponding to radiating a single gluon reads
\cite{kst2,k3p},
\beqn
\sigma^{hN}_1 &=&
\int\limits_0^1 d\alpha_q \int d^2R\,\,
\Bigl|\Psi^h_{\bar qq}
(R,\alpha_q)\Bigr|^2
\int\limits_{\alpha_G\ll 1}
\frac{d\alpha_G}{\alpha_G} 
\int d^2r\nonumber\\ & \times & \, {9\over4}\,\,
\biggl\{\Bigl|\Psi_{\bar qG}(\vec R 
+\vec r,\alpha_G)\Bigr|^2
\sigma_{\bar qq}^N(\vec R +\vec r) +
\Bigl|\Psi_{qG}(\vec r,\alpha_G)\Bigr|^2
\sigma_{\bar qq}^N(r) 
\nonumber\\ & - & \, 
{\rm Re}\,\Psi_{qG}^*(\vec r,\alpha_G)\,
\Psi_{\bar qG}(\vec R +\vec r,\alpha_G)\,
\Bigl[\sigma_{\bar qq}^N(\vec R +\vec r)+\sigma_{\bar qq}^N(r)-
\sigma_{\bar qq}^N(R)\Bigr]
\biggr\}
\label{6.5}
\eeqn
 Here $\alpha_G$ is the fraction of the hadron momentum carried by
the gluon which is assumed to be small;
the notations for the radii are obvious from Fig.~\ref{qqG}.
The first and the second terms in the curly brackets correspond to the
emission of the gluon from the quark and the antiquark respectively,
and the third term to the interference between them.

The nonperturbative wave function for a quark-gluon Fock
component was derived in \cite{kst2}. Neglecting the
quark mass the wave function reads,
\beq
\Psi_{qG}(\vec r,\alpha_G)\Bigr|_{\alpha\ll 1}=
-\,\frac{2\,i}{\pi}\,\sqrt{\alpha_s\over3}\,\,
\frac{\vec e\,^*\cdot\vec r}{r^2}\,
\exp\left(-\,\frac{r^2}{2\,r_0^2}\right)\ ,
\label{6.6}
\eeq
 where $\vec e$ is the polarization vector of the massless gluon.
The mean separation $r_0=0.3\,fm$ is related to 
the nonperturbative light-cone potential
describing the quark gluon interaction. It is fixed 
by the data on large mass diffractive
dissociation corresponding to the triple-Pomeron limit.
The mean  quark - gluon separation 
$r_0$ is much smaller than the distance $R$  
between the quarks. Therefore, one of the $qG$ wave functions in
(\ref{6.5}) can be neglected leaving a factor 2 (a factor 3 in the
case of $NN$ scattering) since both $q$ and $\bar q$ can radiate
the gluon.

At small separations, $r\sim r_0$, the dipole cross section $\sigma_{\bar
qq}^N(r)$ can be evaluated perturbatively and the approximation
$\sigma_{\bar qq}^N(r) = C\,r^2$ can be used. The two-gluon approximation
gives for the factor $C\approx 2.3$ using an effective gluon mass
$m_G=0.15\,GeV$ to incorporate confinement, and $\alpha_s=0.4$ (see below).

Thus, the contribution of the $|\bar qqG\ra$ Fock component to
the total cross section summed over polarization of the radiated
gluon takes the form,
 \beq
\sigma^{hN}_1 = \frac{4\,\alpha_s}{3\,\pi}\,
\ln\left({s\over s_0}\right)\,\,
\frac{9}{4}\,C\,r_0^2\ .
\label{6.7}
\eeq
 Here $\ln[s/s_0]=\ln[(\alpha_G)_{max}/(\alpha_G)_{min}]$
originates from the integration over $\alpha_G$ in (\ref{6.5}), where
$(\alpha_G)_{min}=2/sr_0^2 \approx (1\,GeV^2)/s$, but
$(\alpha_G)_{max}$ is ill defined. It should be small enough,
say $\sim 0.1$ to make sure that the quark-gluon wave functions in
(\ref{6.5}) is independent of $\alpha_G$. Then, assuming that
the quark carries a fraction one third of the proton momentum, we
estimate the value $s_0 \sim 30\,GeV$, which we use in what
follows.

\begin{figure}[tbh] 
\includegraphics{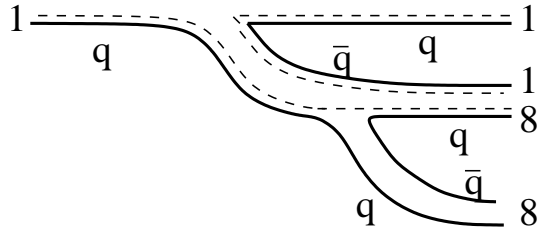} 
\begin{center}
\vspace{3.5cm} 
\parbox{13cm} 
{\caption[Delta] {\sl Radiation of
the second gluon in the leading-log(s)  approximation as seen in
$1/N_c$ approximation when each gluon is replaced by a $\bar qq$
pair. Solid quark lines correspond to the final state, dashed --
lines correspond to the initial state configurations.} 
\label{nG}} 
\end{center}
\end{figure}

The radiation of each new, $n$-th, gluon can be treated as the radiation
by an effective quark, which is the valence quark surrounded by $n-1$
gluons. It should be resolved by the soft interaction with the target as
being different from the radiation of $n-1$ gluons. Therefore, it provides
the same mean cross section $9Cr_0^2/4$ and the factor $4\alpha_s/3\pi$ as
in (\ref{6.7}). This is illustrated in Fig.~\ref{nG} in the $1/N_c$
approximation, {\it i.e.} replacing each gluon by a $\bar qq$ pair.
 According to the general prescription \cite{hir,kst2} of the light-cone
approach, the radiation cross section in the impact parameter
representation is proportional to the total cross section of a colorless
system made of all the final state partons (solid lines in Fig.~\ref{nG})
plus the initial state partons replaced by antipartons (dashed lines).
Since the radiation of
gluons with $\alpha_G \ll 1$ does not affect the impact parameter of the
radiating quark, all the solid and the corresponding dashed lines have the
same impact parameters. Therefore, those quark-antiquark (solid-dashed)
pairs which are color neutral do not contribute to the cross section.
The only quark contributing is the one that radiates the last gluon and
that changes color in a color-octet state with the antiquark as shown
in Fig.~\ref{nG}. Thus, the total
cross section of the multiquark configuration is reduced to one for the
octet-octet dipole with mean separation $r_0$, {\it i.e.} $9\,C\,r_0^2/4$.

In conclusion, the $n$-th term in (\ref{6.3a}) reads
(for a single valence quark), 
 \beq
\sigma^{qN}_n = \frac{1}{n!}\,
\left[\frac{4\,\alpha_s}{3\,\pi}\,\, 
{\rm ln}\left({s\over s_0}\right)\right]^n\,
\frac{9}{4}\,C\,r_0^2\ . 
\label{6.7a} 
\eeq
 Summing up the powers of logarithms in (\ref{6.3a}) we arrive at
the following expression for the total cross section,
 \beq 
\sigma^{pp}_{tot}= \tilde\sigma^{pp}_0 +
3\,\frac{9}{4}\,C\,r_0^2\,\, 
\left({s\over s_0}\right)^{\Delta}\ , 
\label{6.8} 
\eeq 
 where 
 \beq \Delta=\frac{4\,\alpha_s}{3\,\pi}\ . 
\label{6.9} 
\eeq 
\beq \tilde\sigma^{pp}_0=
\sigma^{pp}_0 - \frac{9}{4}\,C\,r_0^2\ .
\label{6.9a}
\eeq      
 Since each of three valence quarks can radiate (see (\ref{6.5}))
the second term in (\ref{6.8}) acquires factor $3$.

The structure of Eq.~(\ref{6.8}) reflects the physical input illustrated in
Fig.~\ref{3q}, as discussed in the Introduction (section~1.2). The energy
dependence of the cross section is related to the excitation of the small
spots (constituent quarks) inside the hadron, while a large energy
independent contribution corresponds to the soft interaction of the
valence quarks skeleton of the hadrons leading to no gluon radiation.
These two parts of the cross section cannot match to provide a power
dependence, $s^{\Delta}$, as a common factor, as it is usually assumed to
the the case
in the so-called soft Pomeron approach \cite{dl,kl,kkl}. This is the
origin of the more complicated form we find {\it i.e.} of the structure
of eq. (\ref{6.8}) or, which is the same, of eq. (\ref{stot}).
The energy dependent part of the cross section, {\it i.e.} the second term
in (\ref{6.8}), is suppressed by the smallness of $r_0^2$ and is expected
to be relatively small at medium-high energies, but grows more steeply 
with energy than the overall $\sigma^{pp}_{tot}$. Note that such a
structure of the total cross section was also suggested in \cite{knp},
however with a different physical motivation. It was found to
fit well the data on total cross section (see below).

The power $\Delta$ in (\ref{6.8}) is related to the Pomeron intercept,
$\alpha_P(0)=1+\Delta$, and can be predicted using (\ref{6.9}) provided
the QCD coupling $\alpha_s$ at virtuality $\sim 1/r_0$ is known.  In
Gribov's theory of confinement \cite{gribov,ewerz} the radius of a
constituent quark is at the borderline between the perturbative and the
nonperturbative QCD regimes. At larger distances, the chiral symmetry
breaks down and pseudo-Golstone pions emerge\footnote{The Pomeron
properties at large distances related to pion loops have been employed in
the model suggested in \cite{kmr}}. At smaller distances, perturbative
QCD is at work. The corresponding critical value of $\alpha_s$ is
\cite{gribov,ewerz},
 \beq
\alpha_c = \frac{3\,\pi}{4}\left(1-\sqrt{{2\over3}}\right)
\approx 0.43\ .
\label{6.10}
\eeq

Another way to evaluate $\alpha_s$ is to average the running QCD coupling
weighted with the transverse momentum distribution of radiated gluons,
\beq 
\la\alpha_s\ra = \frac{\int_0^{\infty} dk_T^2\,
\alpha_s(k_T^2)\,\frac{d\sigma(qN\to qGX)}
{d(ln\alpha_G)\,dk_T^2}} 
{\int_0^{\infty} dk_T^2\,
\frac{d\sigma(qN\to qGX)}
{d(ln\alpha_G)\,dk_T^2}}\ ,
\label{mean-alpha}
\eeq
 where the transverse momentum distribution of gluons radiated in
quark-nucleon interaction is given by Eq.~(130) of \cite{kst2}.  

The standard phenomenological way to extend $\alpha_s(k_T)$ down to small
values $k_T\to 0$ is to make a shift in the argument, $k_T^2 \Rightarrow
k_T^2 + k_0^2$. The value $k_0^2\approx 0.25\,GeV^2$ was estimated in
\cite{ewerz} using dispersion techniques \cite{dmw} of higher twist
effects in hard reactions.

We evaluated (\ref{mean-alpha}) using the dipole cross section
$\sigma_{\bar qq}(\rho)\propto 1-exp(-\rho^2/\rho_0^2)$ which is
proportional to $\rho^2$ at small $\rho$, but levels off at large
separations. The nonperturbative quark-gluon interaction taken into
account in \cite{kst2} is very important, since it squeezes the
quark-gluon fluctuations down to a mean size $r_T\sim r_0$ substantially
increasing the mean transverse momenta of radiated gluons.
Correspondingly, the mean value $\la\alpha_s\ra$ turns out to be rather
small. For the parameter $\rho_0$ varying within a reasonable interval
$0.3 < \rho_0 < 1\,fm$ we found the mean coupling varying between
$\la\alpha_s\ra = 0.38 - 0.43$ which agrees well with the critical value
Eq.~(\ref{6.10}). Substituting the central value of $\la\alpha_s\ra$ into
(\ref{6.9}) and using the interval as the uncertainty for $\alpha_s$ we
get,
 \beq
\Delta = 0.17 \pm 0.01\ .
\label{6.11}
\eeq

This value is about twice as large as the value $0.08$ usually believed to
be required by data. This value, however, can not be compared directly
with the soft Pomeron intercept where it is assumed that the whole cross
section is proportional to $ s^{\Delta}$. The value of (\ref{6.11}) is
just the second term of (\ref{6.8}) and the overall energy dependence is
much less steep as a consequence of the large value of the constant term
$\tilde\sigma^{pp}_0$. We will show later (see Section~\ref{forward}) that
the predicted energy dependence of the cross section (\ref{6.8}) is in a
good accord with the data and corresponds to an effective value
$\Delta_{eff}\approx 0.1$. First of all, however, we should take care of
unitarity since the cross section (\ref{6.8}) violates the Froissart bound
at large $s$.

\section{Impact parameter representation, unitarization}
\label{unitarization}

Although the total cross section stays well below the Froissart-Martin
bound up to the present highest energies, the partial amplitude at small
impact parameters demonstrates a precocious onset of the unitarity
restrictions which are already important in the energy range of existing 
accelerators.

The imaginary part of the partial amplitude corresponding to the total
cross section (\ref{6.3a}) can be decomposed into the terms related by
unitarity to the radiation of different number $n$ of gluons,
 \beq
{\rm Im}\,\gamma_P(s,b)=
\sum\limits_{n}{\rm Im}\,\gamma_n(s,b)\ ,
\label{6.12}
\eeq
where $\gamma_n(s,b)$ is the partial elastic amplitude which depends on
the energy and on the impact parameter $b$. Upon performing the Fourier
transform of the $t$-dependent elastic amplitude, the integral over
$\vec b$ gives the corresponding term $\sigma^{hN}_n$ in (\ref{6.3a}). We
assume that the $t$-dependence of the lowest Fock component which is
related to the spatial distribution of valence quarks is given by the
product of the electromagnetic formfactors of the colliding hadrons (we
confine 
our considerations to $pp$ and $\bar pp$ collisions)  $F^2_p(t)$. For
simplicity we use the standard dipole form for the proton formfactor
$F_p(t) = (1-t\,\la r_{ch}^2\ra/12)^{-2}$, where $\la r_{ch}^2\ra$ is the
mean charge radius squared related to the slope of elastic scattering of
the valence quark skeleton by $B_0=2\,\la r_{ch}^2\ra/3$.

We keep the same $t$-dependence for higher Fock components in (\ref{6.3a})
corresponding to gluon radiation by a projectile valence quark interacting
with the target proton; the slope of these components, however, should
increase linearly with the number of radiated gluons due to their random
walk in the impact parameter plane with a step $\sim r_0^2$ for the
radiation of every new gluon,
 \beq
B_n = {2\over3}\,\left\la r_{ch}^2\right\ra +
\frac{n\,r_0^2}{2}\ .
\label{6.13}
\eeq

The Fourier transform of the square of the dipole formfactor leads to the
following shape for the partial amplitudes \cite{hp},
 \beq
{\rm Im}\,\gamma^{pp}_n(b,s)=
\frac{\sigma^{hN}_n(s)}{8\,\pi\,B_n}\,
y^3\,K_3(y)\ ,
\label{6.14}
\eeq
where $y^2=(4b^2/B_n)^3$, $K_3(y)$ is the third order modified
Bessel function and $\sigma^{hN}_n$ are given by (\ref{6.4}) and
(\ref{6.7a}). The normalization of the partial amplitude is fixed
by the relation, $\sigma_{tot}= 2\,\int d^2b\,{\rm Im}\gamma(b,s)$.

The partial amplitude (\ref{6.12})  rises with energy and eventually
would lead to a violation of the unitarity bound ${\rm Im}\gamma(s,b)\leq
1$, unless unitarity corrections are introduced. Unfortunately, this is
not a well defined procedure since different recipes can be found in the
literature.

The simplest known way to restore unitarity is to eikonalize the partial
amplitude (\ref{6.12}),
\beq
{\rm Im}\,\Gamma_P(b,s)=
1 - {\rm exp}
\Bigl[-{\rm Im}\,\gamma_P(b,s)\Bigr]\ .
\label{6.16a}
\eeq
 At very high $s$ this amplitude approaches the black disk limit
\cite{dklt}, ${\rm Im}\,\Gamma_P(s,b) \to \Theta\bigl[R^2(s)-b^2\bigr]$,
with radius, $R(s)=r_0\,\Delta\,{\rm ln}(s/s_0)$. Correspondingly, at high
energies
 \beq
\Delta\,{\rm ln}\left({s\over s_0}\right) \gg
\frac{\la r_{ch}^2\ra}{r_0^2}
\label{6.17}
\eeq
 all hadronic cross sections reach the maximal universal energy growth
allowed by Froissart-Martin's bound,
 \beq
\sigma^{hN}_{tot}(s) \to 
2\,\pi\,r_0^2\,\Delta^2\,
{\rm ln}^2\left({s\over s_0}\right)\ .
\label{6.18}
\eeq

The eikonalization procedure (\ref{6.16a}) would be suitable if the 
the incoming hadrons were eigenstates of the interaction \cite{zkl}.
Hadrons, however, are subject to diffractive off-diagonal excitation,
and the eikonal form of unitarization should be corrected in a way similar
to Gribov's inelastic corrections \cite{gribov69} for hadron--nucleus
cross sections. The lowest order unitarity correction in (\ref{6.16a})
comes from the quadratic term in the exponent expansion of $\Gamma(b,s)$.
It has to be modified using the AGK cutting rules \cite{agk} to include
single diffraction,
 \beq 
{\rm Im}\,\Gamma_P = 
{\rm Im}\,\gamma_P - 
{1\over2}\,\Bigl({\rm Im}\,\gamma_P\Bigr)^2\,
\Bigl[1\,+\,D(s)\Bigr]\,+\, 
O\bigl(\gamma_P^3\bigr)\ ,
\label{6.19} 
\eeq 
where $D(s) =\sigma_{sd}(s)/\sigma_{el}(s)$ is approximately $0.25$ in the
ISR energy range and decreases slightly with energy
$\propto s^{-0.04}$ \cite{dino,schlein}. Indeed,
$\sigma_{el}(s)=\sigma_{tot}^2(s)/(16\pi B_{el}) \propto
s^{0.1}$, but the energy dependence of the diffractive cross section
is rather flat (due to stronger unitarity corrections), $\sigma_{sd}(s)
\propto s^{0.06}$. Asymptotically, as $s\to\infty$, $D(s)$ vanishes since
$\sigma_{el}(s) \propto {\rm ln}^2s$ and $\sigma_{sd}(s) \propto {\rm
ln}s$.

The inelastic corrections to higher order terms in the expansion
(\ref{6.16a}) are poorly known. A simple way to keep (\ref{6.19}) and to
include diffraction into the higher terms is to modify (\ref{6.16a}) as,
 \beq
{\rm Im}\,\Gamma_P(b,s)=
\frac{1}{1+D(s)}\,\left\{
1 - {\rm exp}
\left[-\Bigl(1+D(s)\Bigr)\,
{\rm Im}\,\gamma_P(b,s)\right]
\right\}\ ,
\label{6.20}
\eeq
which is known as quasi-eikonal model \cite{kaidalov}.

A more consistent way of unitarization suggested in \cite{knp} prescribes
to use the eikonal expression (\ref{6.16a}) in terms of the color dipole
cross sections and then average it over the transverse separations of all
partons. Unfortunately, this procedure is simple only if the dipole cross
sections depend quadratically on the separation parameter, which is
definitely incorrect for $\sigma_{\bar qq}^N(R)$ in (\ref{6.4}). If it
were true one would have $D(s)=1$ which would exceed four times the
experimental value.  We have tried this unitarization prescription as well
and found that the results still agree with data pretty well.
Nevertheless, we use for further applications Eq.~(\ref{6.20}) since it
explicitly exploits experimental information, correctly reproduces the
lowest unitarity correction (\ref{6.19}), and is rather accurate within
the energy range of interest.

\section{\bf Comparison with forward scattering data}
\label{forward}

All the parameters in Eqs.~(\ref{6.7}), (\ref{6.14}) and (\ref{6.20}) are
known, except the Born cross section $\tilde\sigma_0$. Although it may be
estimated in various models, none of these is sufficiently reliable and we
choose to determine it so as to best reproduce the data. As soon as the
absolute normalization of the total cross section is fitted at some energy
and the parameters $\tilde\sigma_0$ is fixed, the energy dependence can be
predicted. For this comparison we selected 
the data \cite{cdf} for $\sigma^{\bar pp}_{tot}$ at $\sqrt{s}=546\,GeV$ as
being the most precise.
In addition, this energy value is high enough that we neglect the Reggeon
contribution. We calculate $\sigma^{\bar pp}_{tot}$ from 
 \beq 
\sigma_{tot}= 2\,\int d^2b\,{\rm Im}\,\Gamma(b,s)\ ,
\label{6.15a}
\eeq
using (\ref{6.20}) and fix our only unknown parameter at
$\tilde\sigma_0=39.7\,mb$.

\begin{figure}[thb]
\includegraphics{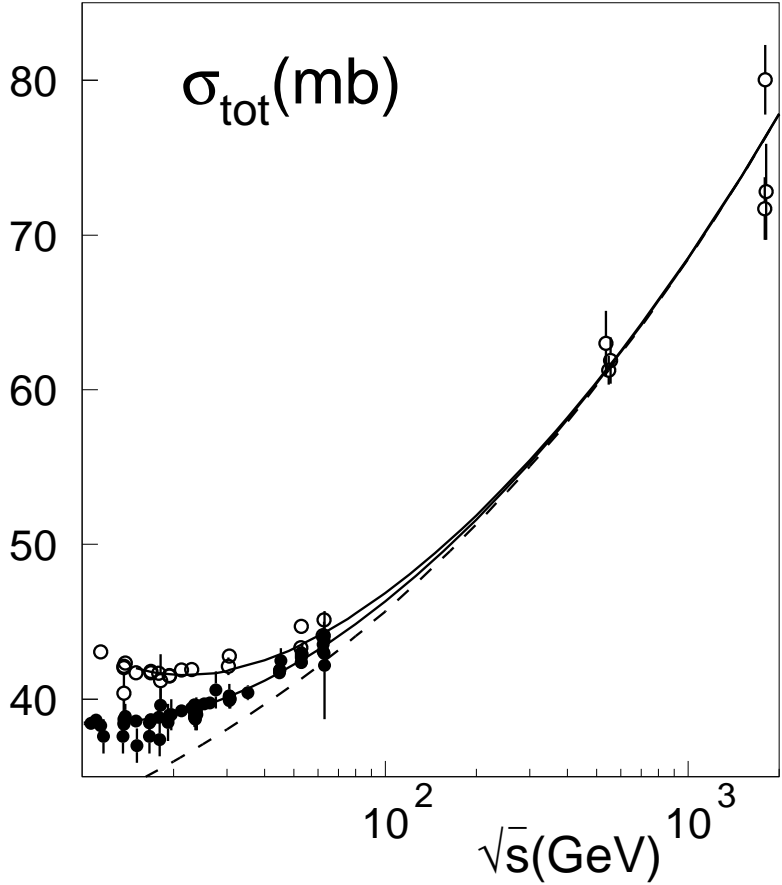}
\begin{center}
\vspace{12cm}
\parbox{13cm}
{\caption[shad1]
 {\sl Data for total $pp$ (closed circles) and $\bar pp)$ (open
circles) cross sections \cite{pdt1} at $\sqrt{s} > 10\,GeV$.
The dashed curve shows the predicted energy dependence for the
net Pomeron contribution whose normalization is fixed by the 
$\sqrt{s}=546\,GeV$ data \cite{cdf}. The solid curves, (bottom
for $pp$ and upper for $\bar pp$), represent the results corrected for
Reggeon contribution which is fitted to the data.}
\label{sigtot}}
\end{center}
\end{figure}

Now we are in the position to predict the energy dependence for the
total cross sections and compare it with $pp$ and $\bar pp$ data.
The result is depicted by the dashed curve in Fig.~\ref{sigtot}
is in a good agreement at high energies, but somewhat off the data at
medium high energies. This is not surprising since the (secondary) Reggeon
contribution is still missing and this is well known to be important at
medium high energies.

In order to improve the description of the data one should, therefore add
the contribution of leading Reggeons with intercept $\alpha_R(0)\approx
1/2$. This should be done directly in the partial elastic amplitude,
 \beq
{\rm Im}\,\Gamma(s,b) = {\rm Im}\,\Gamma_P(s,b) + 
\Gamma_R(s,b)\,\Bigl[1\,-\,{\rm Im}\,\Gamma_P(s,b)\Bigr] .
\label{6.15b}
\eeq
The Reggeon term is suppressed by the absorptive corrections which have
the same origin as those which slow down the energy dependence of the
diffraction cross section mentioned above. The Reggeon term is
parameterized as,
 \beq {\rm Im}\,\Gamma_R(s,b) =
\frac{\sigma_R}{4\,\pi\,B_R(s)}\,
\left(\frac{s}{s_0}\right)^{\alpha_R(0)-1}\, 
{\rm exp}\left(-\,\frac{b^2}{2\,B_R(s)}\right)\ , 
\label{6.16}
\eeq
and $B_R=R_R^2+2\,\alpha_R^{\prime}\,{\rm ln}\,s$. We fixed the standard
values the parameters $\alpha_R(0)=0.5$ and
$\alpha^{\prime}_R=0.9\,GeV^{-2}$, but fitted $R_R^2=3\,GeV^{-2}$. We also
fitted the normalization factors which are very different for $pp$ and
$\bar pp$ because of the (approximate) exchange degeneracy. We found
$\sigma_R^{pp}= 17.8\,mb$ and $\sigma_R^{\bar pp}= 32.8\,mb$. The result
of the fit, shown in Fig.~\ref{sigtot} by solid curves demonstrates a
good agreement with data. As anticipated, the Reggeon corrections are
important only in the ISR energy range and below.

\begin{figure}[thb]
\includegraphics{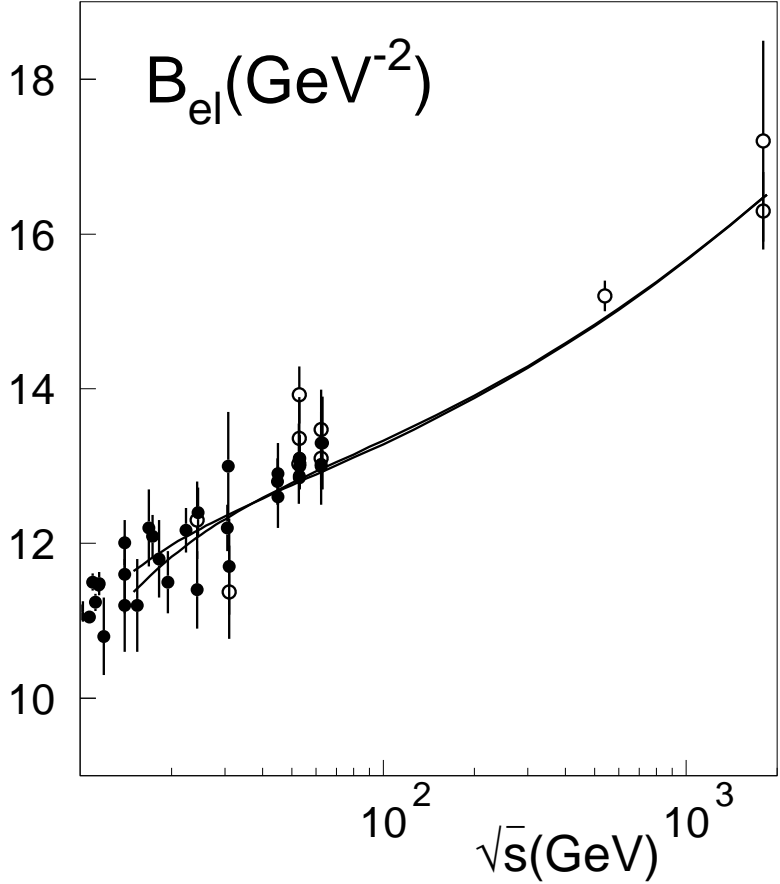}
\begin{center}
\vspace{12cm}
\parbox{13cm}
{\caption[shad1]
 {\sl Data for the elastic slope \cite{bel} and our predictions.
The upper and bottom curves and, correspondingly, the open and full
circles belong to $\bar pp$ and $pp$, respectively.}
\label{slope}} 
\end{center} 
\end{figure}

Since $\tilde\sigma_0$ and all the Reggeon parameters are now fixed,
we can predict the slopes of elastic $pp$ and $\bar pp$ scattering (both
the absolute values and energy dependence). We calculate the slope using
the relation,
 \beq B_{el}(s) \,=\,{1\over2}\,\la b^2\ra\,=\,
\frac{1}{\sigma_{tot}}\,
\int d^2b\, b^2\,{\rm Im}\,\Gamma(b,s)\ .
\label{6.15} 
\eeq 
Once again, the results shown in Fig.~\ref{slope} in comparison with
data for $pp$ and $\bar pp$ scattering demonstrate a good agreement.
Although we had some freedom in the choice of the proton formfactor and of
the proton charge radius, this affects only the absolute value of the
slope. The energy dependence is fully predicted. Since it describes the
data well, we correctly predict the effective Pomeron slope
$\alpha^{\prime}_{eff}\approx 0.25\,GeV^{-2}$.

Note that, often, phenomenological fits treat total cross sections and
slopes as controlled by different parameters. In these cases, one cannot
predict the energy dependent slope even if the total cross section is
known. 

The radiation of every new gluon leads to an expansion of the gluon cloud
by a ``step'' $\delta\la r^2\ra \approx (0.3\,fm)^2$.  Eventually, the
initial approximation of a small gluon cloud inside a large hadron will
break down. This, however, will happen only at very high energies. The
mean number of gluons in a quark $\la n\ra = \Delta\,{\rm ln}(s/s_0)$ is
quite small, $\la n\ra = 0.5-0.8$ at ISR, about $\la n\ra \approx 1.5$ at
$S\bar ppS$ and reaches $\la n\ra \approx 2$ at the Tevatron. Therefore
the
mean radius of a constituent quark is still rather small and our
approximation remains quite valid and we should expect it to
break down only at very high energies which are well beyond the range of
present accelerators. Of course the radius of an constituent quark, {\it
i.e.} the radius of the gluon cloud depends also on a reference frame,
{\it e.g.} in the c.m. it is twice as small as in the rest frame of the
target.

\section{Elastic scattering data analyzed in the impact
parameter}\label{impact}

The partial amplitude (\ref{6.20}) has nontrivial $s$- and
$b$-dependences. It is nearly energy independent for central collisions,
but steeply grows with energy on the periphery as it was first found in
the analysis of the data by Amaldi and Schubert \cite{as}. These
properties are averaged out and hidden in the total or differential
elastic cross sections. To extract information about the shape of the
partial elastic amplitude in the impact parameter representation from the
data on elastic $pp$ and $\bar pp$ scattering, we follow the procedure
used in \cite{as}. However, to make the analysis less model dependent we
fit differential elastic cross section data independently at each energy,
thus, no model for energy dependence is involved. The geometrical scaling
model used in \cite{as} assumes that the total cross section is
proportional to the slope of the elastic differential cross section,
$\sigma^{pp}_{tot}(s) \propto B^{pp}(s)$. This relation is, however, a
result due to the accidental closeness of the Regge model parameters
characterizing the energy dependence of the cross section,
$\sigma^{pp}_{tot}(s)\propto s^{\Delta}\approx 1+\Delta\,{\rm ln}s$ and of
the slope, $B^{pp}(s)=B^{pp}_0 + 2\,\alpha^{\prime}_P\,{\rm ln}s$, where
$\alpha^{\prime}_P\approx 0.25\,GeV^{-2}$ and $B^{pp}_0\approx
7.5\,GeV^{-2}$. Indeed, the effective Pomeron intercept
$\Delta=\alpha_P(0)-1 \approx 0.08$ is close to the ratio
$2\alpha_P^{\prime}/B^{pp}_0\approx 0.067$. Obviously, geometrical scaling
may occur only in a restricted energy range (namely, in the ISR energy
range used in \cite{as}) and it had been predicted \cite{dklt} to break
down at higher energies. This was confirmed later by the $Sp\bar pS$ and
Tevatron data.

Since we are interested in the Pomeron part of the elastic amplitude, the
data for differential cross section of elastic $pp$ and $\bar pp$
scattering selected for the analysis cover the wide range of high energies
including $ISR$\footnote{We rely upon the relative normalizations of
differential cross sections measured in different $t$-intervals at the
same energy given in \cite{as}}, and $Sp\bar pS$ \cite{data}. We do not
include the data from the Tevatron since they are available only in too
narrow a range of $t$ which is not sufficient for Fourier transformation.
We parameterize the imaginary and real parts of the elastic scattering
amplitude in a model independent way as,
 \beq
{\rm Im}\,f(t)=\sum\limits_{i=1}^{3}
a_i\,e^{b_i\,t};
\label{data.1}
\eeq
\beq
{\rm Re}\,f(t)=\sum\limits_{i=1}^{2}c_i\,e^{d_i\,t}
\label{data.2}\ ,
 \eeq where $a_i,\ b_i,\ c_i,\ d_i$ are parameters to be fitted. The
amplitudes are related to the cross sections as,
\beq
\frac{d\,\sigma}{d\,t}= 
\bigl[{\rm Re}\,f(t)\bigr]^2 +
\bigl[{\rm Im}\,f(t)\bigr]^2\ ;
\label{data.3}
\eeq
\beq
\sigma_{tot}=4\,\sqrt{\pi}\,{\rm Im}\,f(0)\ .
\label{data.4}
\eeq

To make the normalization of the differential cross section data more
reliable, first we perform a common fit of the $pp$ and $\bar pp$ total
cross sections with the same Pomeron part as function of energy. Then we
adjust the normalizations of the differential elastic cross section data
to the optical points, {\it i.e.} demand that $4\,\sqrt{\pi}\,\sum
a_i=\sigma_{tot}$ at each energy. The data \cite{rho} for $\rho(s)$ ratio
of the real to the imaginary parts of the amplitudes at $t=0$ were also
involved in the analysis. We fit these data by a smooth energy dependence
and demand then $\sum c_i=\rho\,\sum a_i$ for each energy included in the
analyses of differential cross sections. We performed two variants of fit,
 \begin{itemize}
\item
{\it variant I:} both $c_1$ and $c_2$ are used as free parameters in the
fit; 
\item
{\it variant II:} $c_2=0$ in (\ref{data.2}).
\end{itemize}

\begin{figure}[thb]
\includegraphics{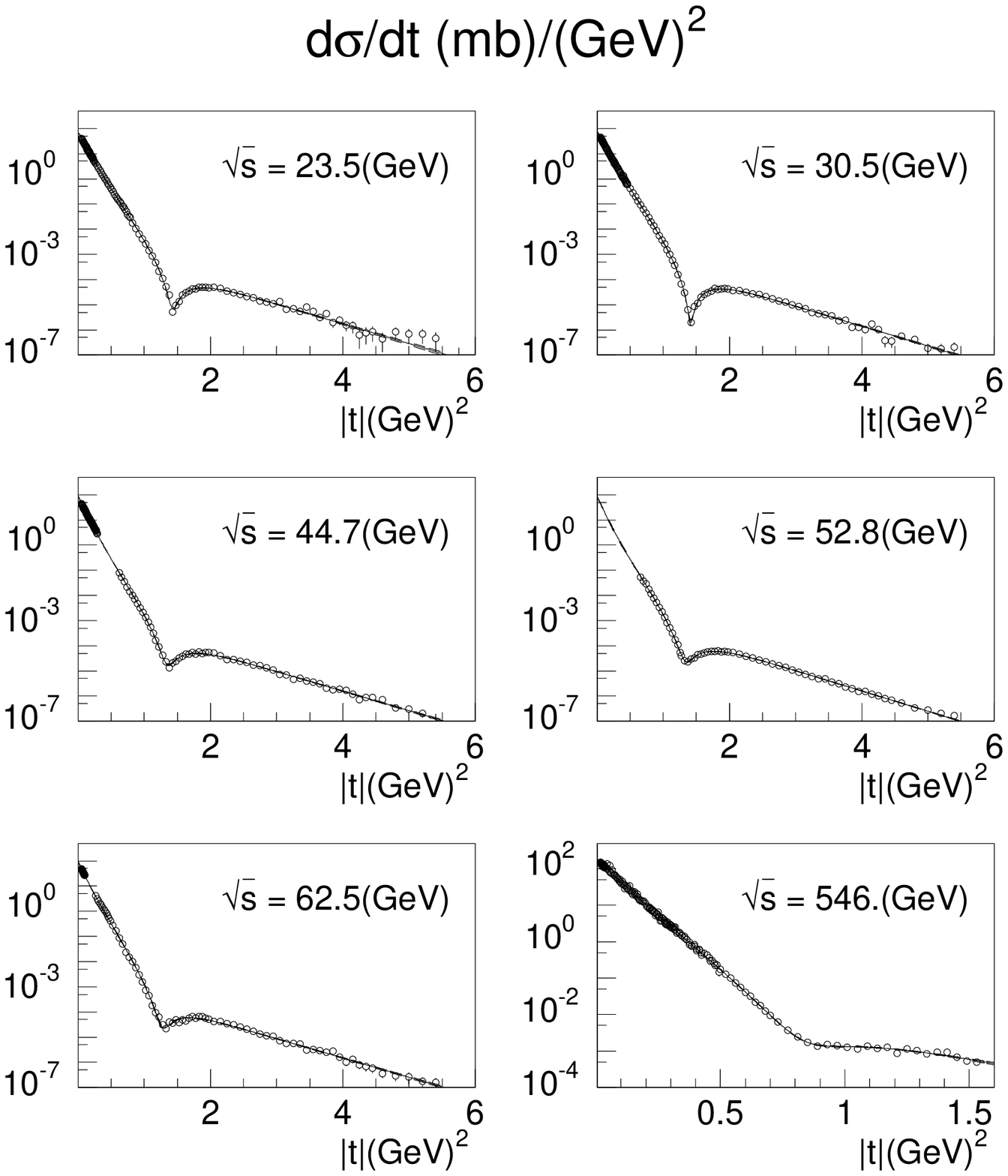}
\begin{center}
\vspace{17cm}
\parbox{13cm}
{\caption[shad1]
{\sl The differential cross sections of elastic $pp$ and $\bar pp$
scattering at different energies.
The first five panels show the $pp$ data from ISR \cite{as}, the last one
the $\bar pp$ data \cite{data} from $Sp\bar pS$.
The curves show our fit Eqs.~(\ref{data.1})-(\ref{data.3}) in
{\it variant I}.}
\label{sdiff}}
\end{center}
\end{figure}

The data in the fit and the results in {\it variant I} are
depicted in Fig.~\ref{sdiff}.

\vspace*{0.1cm}

As soon as the parameters in (\ref{data.1}) and (\ref{data.2}) are found
we can calculate the partial amplitude in the impact parameter
representation at each energy ,
\beq
\Gamma(b)=\frac{1}{2\,\pi^{3/2}}
\int d^2q\,e^{i\,\vec q\cdot\vec b}\,f(-q^2)\ ,
\label{data.5}
\eeq
where $\vec q$ is the transverse component of the momentum transfer,
$t\approx - q^2$. It is normalized according to (\ref{6.15a}).

\begin{figure}[thb]
\includegraphics{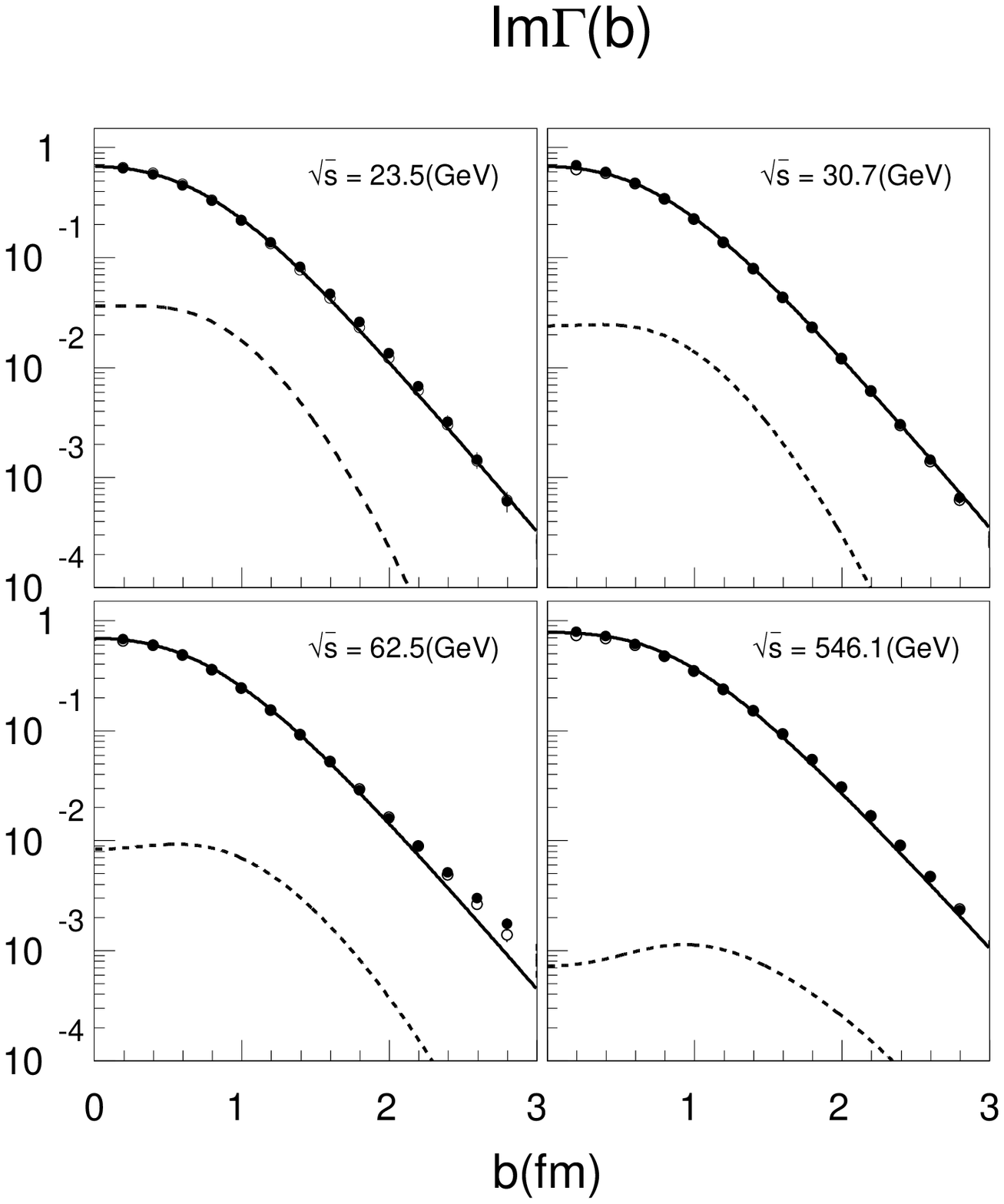}
\begin{center}
\vspace{15cm}
\parbox{13cm}
{\caption[shad1]
{\sl The imaginary part of the partial 
amplitude ${\rm Im}\,\Gamma(b)$ as function of the
impact parameter at different energies. The first three
panels correspond to the ISR, the last one to the $Sp\bar pS$ data.
The curves show our theoretical prediction with Eq.~(\ref{6.20})
using the parameters given in Table~1 and obtained fitting the 
$t=0$ data (total cross sections and slopes).}
\label{gam-b}}
\end{center}
\end{figure}

 A few examples of our results for ${\rm Im}\,\Gamma(b)$
corresponding to {\it variant I} (full points) and {\it
variant II} (open points) are shown in Fig.~\ref{gam-b} 
with spacing $0.2\,fm$ in the impact parameter and for a
few energies.
The errors are calculated using the error matrix resulting from the fit.

One can see that at $b=0$ the amplitude nearly saturates the unitarity
limit and hardly changes with energy, while at larger impact parameters the
amplitude grows quite substantially.

Our predictions including the Pomeron contribution and Reggeon part are
compared with the data in Fig.~\ref{gam-b}. The Reggeons shown by dashed
curves are calculated for $pp$ and $\bar pp$ interactions for ISR and
S$\bar pp$S data respectively. Their contribution is quite a small fraction
of the full partial amplitude represented by the solid curves. The 
agreement between the data and our predictions is remarkably good,
especially if we recall that the Pomeron part has no free parameters,
except one, $\tilde\sigma_0$, adjusted to the total cross section measured
at one energy $\sqrt{s}=546\,GeV$ \cite{cdf}. Both the predicted shape of
the partial amplitude and its energy development are confirmed by the
data.

\section{Pomeron trajectory in the impact parameter
space}\label{trajectory}

\begin{figure}[thb]
\includegraphics{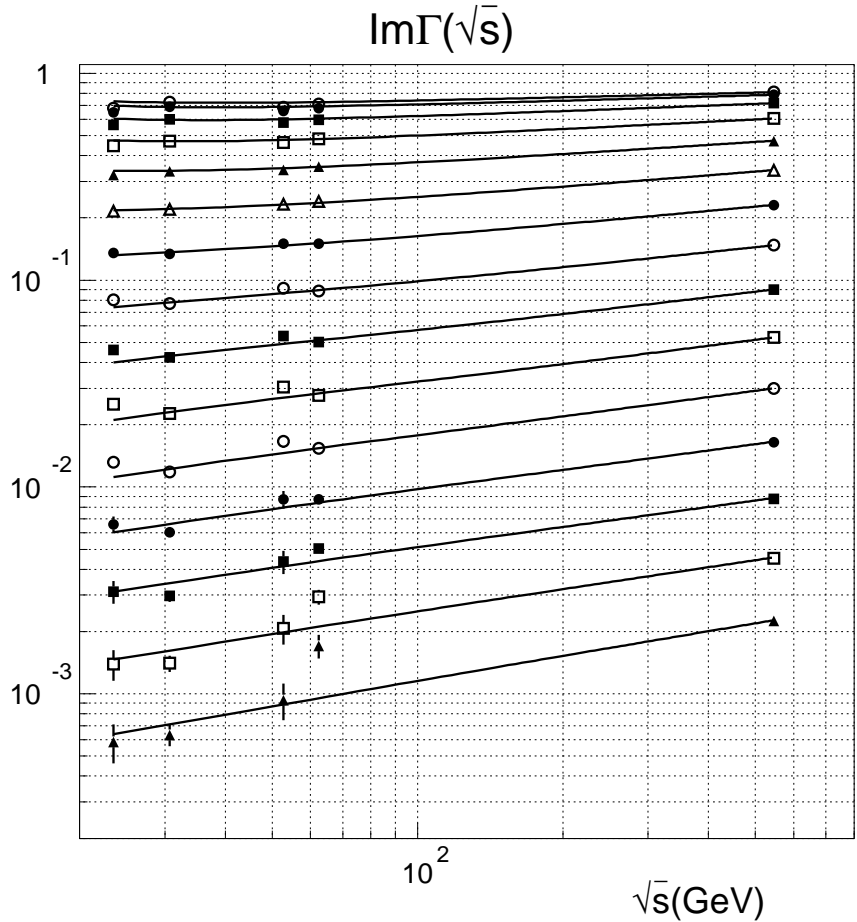}
\begin{center}
\vspace{12cm}
\parbox{13cm}
{\caption[shad1]
 {\sl ${\rm Im}\,\Gamma(b)$ plotted at various values of $b=0.0,\ 0.2,\
0.4,\ ...\ 2.8\,fm$ as function of energy. The values and error bars
correspond to Fig.~\ref{gam-b}. The lines correspond to the fit with power
dependence on energy.}
 \label{gam-s}}
\end{center}
\end{figure}

\begin{figure}[thb]
\includegraphics{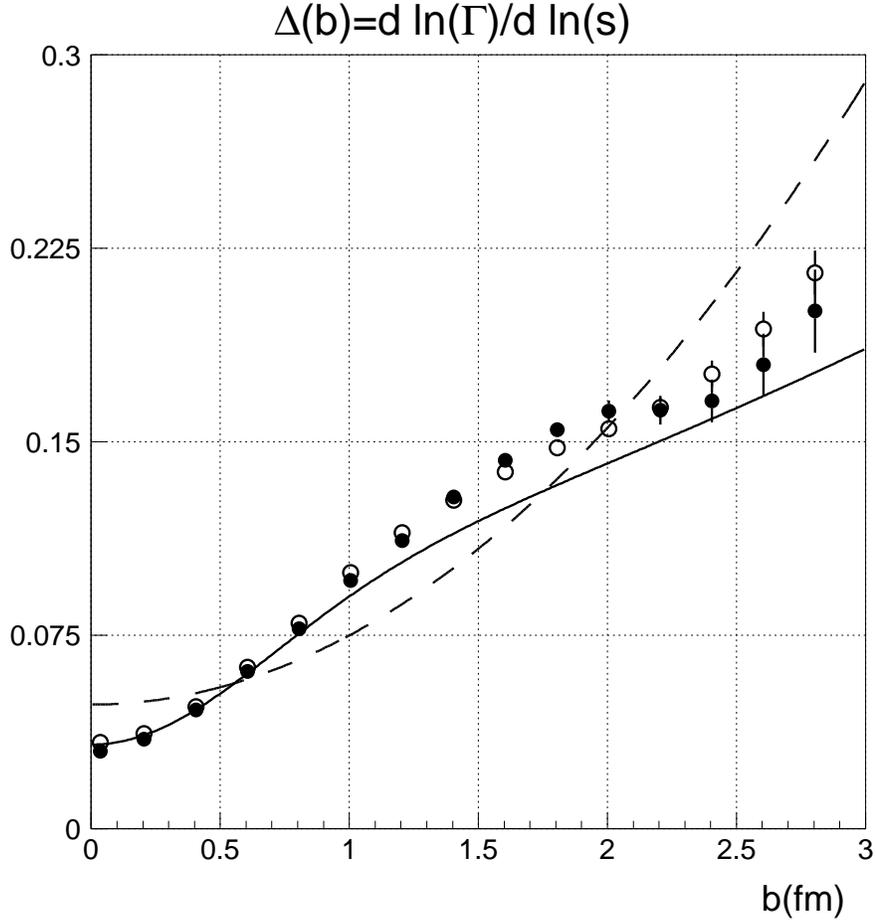}
\begin{center}
\vspace{12.5cm}
\parbox{13cm}
{\caption[shad1]
{\sl The exponent $\Delta(b)$ found by the fit to each point of
Fig.~\ref{gam-s} with power dependence on energy at each value of $b$. The
black and open points correspond to  the fits with parameterizations I and
II respectively.
Our predictions with Eq.~(\ref{data.7}) are shown by solid curve.
The dashed curve demonstrates prediction of a single Regge pole model
without any unitarity corrections.}
\label{delta}}
\end{center}
\end{figure}

The partial elastic amplitude rises with energy faster for peripheral than
for central collisions. The energy dependence of ${\rm Im}\,\Gamma(b,s)$
at different values of the impact parameter is shown in Fig.~\ref{gam-s}
for {\it variant I}.
One can see by the eye that the upper curves corresponding to central
collisions are nearly horizontal, while the bottom ones representing
peripheral collisions rise steeply with energy.  The curves show the
results of the fit to the data for energy dependence of the partial
amplitude at different values of the impact parameter by the expression
\beq
{\rm Im}\,\Gamma(b,s) = \Gamma_0\,s^{\Delta(b)} +
{\rm Im}\,\Gamma_R(b,s)\ ,
\label{data6a}
\eeq
in which the Reggeon term (\ref{6.16}) is calculated with the parameters
fixed by the fit to total cross section data. The exponent $\Delta(b)$
varies with the impact parameter and is fitted to the data for what
concerns its energy dependence in each bin of $b$ as it is shown in
Fig.~\ref{gam-s}. We ignored the data at $\sqrt{s}=62\,GeV$ since they are
too much off the smooth interpolation of the data at lower and higher
energies. The results of the fit for $\Delta(b)$ are plotted in
Fig.~\ref{delta}
 as a series of black points corresponding to each of the lines in
Fig.~\ref{gam-s}. Open points show the results of the fit corresponding to
{\it variant II}. The error bars are determined by the error matrix of the
fit. Note that these values of $\Delta(b)$ correspond to the Pomeron
contribution since the Reggeon part is sorted out.

Although $\Delta(b)$ is very small $\sim 0.03$ for central collisions,
({\it i.e} near $b=0$), it increases dramatically (nearly by one order of
magnitude) at large $b$. Thus, the data show that the energy dependence of
the total cross section originates mainly from peripheral interactions.
This confirms the observation of \cite{as}.

The systematic uncertainty of our analysis is related mainly to the choice
of parameterization for the elastic amplitude. The difference between
variants I and II can be treated as a characteristic estimate of this
uncertainty. There is no significant difference between the two solutions.

Our predictions plotted as solid curves in Fig.~\ref{gam-b} can also be
translated into values for the effective exponent $\Delta(b)$,
 \beq
\Delta_{eff}(b)=\frac{d\,{\rm ln}\Bigl[\Gamma_P(s,b)\Bigr]}
{d\,{\rm ln}s}\ ,
\label{data.7}
\eeq
using the theoretical amplitude (\ref{6.20}) with the same parameters
already determined. The results are shown as a solid curve in
Fig.~\ref{delta}. It agrees well with the data. 

The dashed curve shows the prediction of the simplest parameterization
\cite{dl,pdt} for the elastic amplitude with power $s$ -~ and exponential
$t$ - dependences for both the Pomeron and Reggeon terms.  Although the
unitarity corrections are neglected, this parameterization is indeed quite
successful in describing total cross sections and elastic slopes
\cite{pdt}. One can see, however, that its agreement with the data
in Fig.~\ref{delta} is quite poor. It overshoots the data for central
(lack of unitarization) and peripheral collisions and has quite a
different $b$-dependent shape. Nevertheless, one should not interpret 
the smallness of $\Delta$ at $b=0$ as a manifestation of saturation of
unitarity.

\section{Summary, discussion and outlook}\label{summary}

We present the first successful quantitative dynamical description of
small angle elastic scattering of light hadrons. The key points of our
approach are,
\begin{itemize}

\item
The data for diffractive gluon radiation (large mass diffraction) demand
a small transverse separation $r_0\approx 0.3\,fm$ between the radiated
gluon and the valence quark \cite{kst2,hs}.

\item
A new regime is found which allows explicit calculations: the gluon
clouds of valence quarks are much smaller than the hadronic size,
$r^2_0/R_h^2\ll 1$. This is a different limiting case compared to DIS
where the gluon cloud is much larger than the $\bar qq$ separation.

\item
The interference between amplitudes of gluon radiation by different
quarks is suppressed as ${\rm exp}(-R_h^2/r_0^2)$ leading to an additivity
of the valence quarks in the part of the total cross section related
to gluon radiation. 

\item 
 Since gluon radiation is controlled by the semi-hard scale $Q\sim 2/r_0$
it can be evaluated perturbatively. The radiation cross section is
suppressed by the small factor $r_0^2\approx 1\,mb$, but steeply rises with
energy $\propto s^{\Delta}$, where $\Delta$ is given by Eq.~(\ref{6.11}).

\item 
 The part of the total cross section $\tilde\sigma_0$ related to soft
collisions without excitation of the valence quarks is large since it is
controlled by the large hadronic radius. It is independent of energy and
may cause deviation from quark additivity. 

\item
The two-scale structure ($r_0$ versus $R_h$) of light hadrons unavoidably
leads to the specific form (\ref{stot}) of energy dependence for the total
cross section. The terms with and without gluon radiations are governed by
different scales and cannot match in order to exponentiate into a common
factor $s^{\Delta}$ for the whole cross section.

\item
 While the second, energy dependent term in Eq.~(\ref{stot}) can be
evaluated perturbatively, none of current models can estimate reliably the
first constant term $\tilde\sigma_0$. We treat it as a free parameter and
fix it by the normalization of total cross sections. Then we predict the
energy dependence of the total cross section and the forward elastic slope
in good accord with data.

\item
As a further rigorous test of the model, we perform a model-independent
analysis of available high-energy data for the elastic differential cross
sections and extract the partial amplitudes in impact parameter
representation.  Our model describes well the observed $b$- and $s-$
dependences of the partial amplitude.

\item 
We also extract the Pomeron trajectory $\alpha_P(b)=1+\Delta(b)$ in the
impact parameter representation. $\Delta(b)$ is very small at $b=0$ as a
result of unitarity saturation, but rises by an order of magnitude for
peripheral collisions, in good accord with our predictions. 

\end{itemize}

Concluding, the strong interaction of radiated gluons is vital for
present approach. It squeezes the gluon clouds of valence quarks and
allows to apply perturbative QCD to calculation of the radiation cross
section. It is worth emphasizing that it is not legitimate to mimic these
nonperturbative effects introducing an effective gluon mass $m_G\sim
0.7\,GeV$ as is frequently done in the literature. Indeed, only the
light-cone gluons interact nonperturbatively during their long lifetime.
However, the t-channel Coulomb gluons cannot be treated on the same
footing as the light-cone ones, their lifetime is always short. These
gluons are massless and can propagate far away.  To incorporate the
confinement one should assign only a small effective mass $m_G\sim
\Lambda_{QCD}$ to the t-channel gluons. Making these t-channel gluons as
heavy as those on the light-cone would suppress the factor $C$ in
(\ref{6.8}) and the term $\sigma_1$ in (\ref{stot}) by nearly order of
magnitude in contradiction with data.

Note that our results explain the surprisingly high effective Pomeron
intercept $\Delta\approx 0.2$ observed in diffractive DIS.
It is known that diffraction is dominated by soft interactions even at
high $Q^2$ and one could expect about twice smaller value.
However, large mass diffraction related to diffractive gluon 
radiation corresponds to the second term in Eq.~(\ref{6.8}), {\it i.e.}
value of $\Delta$ given by (\ref{6.11}).

Although data for diffractive dissociation in soft hadronic collisions
were used to fix the strength of the nonperturbative gluon interaction,
one can further test the model performing a similar analysis of data
for the differential cross section of single diffraction in the impact
parameter representation. Such an analysis is in progress and will be
published elsewhere. 

In this paper we concentrate on calculation of the elastic hadronic
amplitude related via unitarity to inelastic processes like gluon
radiation. It is natural to extend the test of the model comparing
directly to data for multiparticle production. In particular, the AGK
cancelation \cite{agk} of unitarity corrections leads to the inclusive
cross section in central region of rapidities rising with energy as
$s^\Delta$ where $\Delta$ has its genuine value not disturbed by
unitarity corrections. Such an analysis of data performed in \cite{lhd}
has led to a surprisingly similar conclusions as ours. Namely, (i) the
data cannot be described by the energy dependence $\propto s^\Delta$ in
the whole energy range, but demand an additional constant term; (ii) the
fit to data resulted in $\Delta=0.17$, exactly what is predicted by our
calculations.

Note that the model suggested in \cite{wang} which describes
multiparticle production in terms of energy independent string
fragmentation and rising with energy mini-jet contribution also goes
along with the basics of our model.

{\bf Acknowledgments:} We are thankful to J\"org H\"ufner, Hans-J\"urgen
Pirner, Andreas Sch\"afer and Sasha Tarasov for inspiring and helpful
discussions. We are grateful to Anatoli Likhoded who informed us about
the analysis of data for inclusive cross sections performed in
Ref.~\cite{lhd} which perfectly agree with our results.  This work was
partially supported by the grant INTAS-97-OPEN-31696, by the European
Network: Hadronic Physics with Electromagnetic Probes, Contract No.
FMRX-CT96-0008 and by the INFN and MURST of Italy.


\begin{thebibliography}{MMM}

\bibitem{k3p}  B.Z.~Kopeliovich, I.K.~Potashnikova, B.~Povh and
E.~Predazzi, Phys. Rev. Lett. {\bf 85} (2000) 507.

\bibitem{dklt} M.I.~Dubovikov, B.Z.~Kopeliovich, 
L.I.~Lapidus and K.A.~Ter-Martirosyan, Nucl. Phys.
{\bf B123} (1977) 147.

\bibitem{agk} A.V.~Abramovsky, V.N.~Gribov
and O.V.~Kancheli, Yad. Fiz. {\bf 18} (1973) 595.

\bibitem{ck} A.~Capella, A.~Kaidalov, C.~Merino and J.~Tran~Thanh~Van,
Phys. Lett. {\bf B337} (1994) 358.

\bibitem{bgp} M.~Bertini, M.~Giffon and E.~Predazzi, Phys.
Letters {\bf B349} (1995) 561.

\bibitem{dgmp} P.~Desgrolard, M.~Giffon, E.~Martynov and E.~Predazzi,
DFTT-66-99 (Dec. 1999), Hep-ph 0001149 (to appear on Eur. Phys. J.).

\bibitem{dl} A.~Donnachie and P.~Landshoff, Nucl. Phys.
{\bf B303} (1988) 634.

\bibitem{pdt} Review of Particle Properties, Phys. Rev. {\bf D54} (1996) 1.

\bibitem{bfkl} E.A.~Kuraev,
 L.N.~Lipatov and V.S.~Fadin, Sov.  Phys.  JETP
 {\bf 44} (1976) 443 ; {\bf 45} (1977) 199;
 Ya.Ya.~Balitskii and L.I.~Lipatov, Sov.  J.  Nucl.
 Phys. {\bf 28} (1978) 822; L.N.~Lipatov,
 Sov. Phys. JETP {\bf 63} (1986) 904.

\bibitem{book} Yu.L.~Dokshitzer, V.A.~Khoze,
 A.H.~Mueller and S.I.~Troyan, {\sl Basics of
 Perturbative QCD}, Editions Frontieres, ADAGP,
 Paris 1991.

\bibitem{review} B.~Badelek, K.~Charchula, M.~Kravczyk
and J.~Kwiecinski, Rev. Mod. Phys. {\bf 64} (1992) 927.

\bibitem{kst2} B.Z.~Kopeliovich, A.~Sch\"afer and A.V.~Tarasov,
Phys. Rev. {\bf D62} (2000) 054022 (hep-ph/9908245).

\bibitem{hs} F.~Hautmann and D.E.~Soper, {\sl Color transparency in deeply
inelastic diffraction}, hep-ph/0008224

\bibitem{braun} V.M.~Braun, P.~G\'ornicki, l.~Mankiewicz
and A.~Sch\"afer, Phys. Lett. {\bf B302} (1993) 291.

\bibitem{pisa} M.~D'Elia, A.~Di~Giacomo and E.~Meggiolaro, 
Phys. Lett. {\bf B408}, 315 (1997).

\bibitem{shuryak1} T.~Sch\"afer, E.V.~Shuryak, Rev. Mod. Phys. 
{\bf 70}, 323 (1998).

\bibitem{shuryak2}  E.V.~Shuryak, 
{\sl Toward the non-perturbative description of high energy processes},
hep-ph/0001189.


\bibitem{l} F.E.~Low, Phys. Rev. {\bf D12} (1975) 163.

\bibitem{n} S.~Nussinov, Phys. Rev. Lett. {\bf 34} (1975) 1986.

\bibitem{gs} J.F.~Gunion and D.E.~Soper, Phys. Rev. {\bf D15} (1977) 2617.

\bibitem{kl} D.~Kharzeev and E.M.~Levin, Nucl.Phys. {\bf B578} (2000) 351.

\bibitem{kkl} D.~Kharzeev, Yu.~Kovchegov and E.M.~Levin,  
{\sl QCD Instantons and the Soft Pomeron}, hep-ph/0007182.

\bibitem{gn} E.~Gotsman and S.~Nussinov, Phys. Rev. {\bf D 22} (1980) 624.

\bibitem{dosch} H.G.~Dosch, Phys. Lett. {\bf B190}, 177 (1987);
 H.G.~Dosch and Yu.A.~Simonov, Phys. Lett. {\bf B205}, 339 (1988).

\bibitem{pirner}  H.G.~Dosch, T.~Gousset, G.~Kulzinger and H.J.~Pirner,
 Phys. Rev. {\bf D55}, 2602 (1997) 

\bibitem{hp} J.~H\"ufner and B.~Povh, Phys. Rev. {\bf D46} (1992) 990.

\bibitem{p} B.~Povh, {\sl Hadron Interactions - Hadron Sizes},
Advances in Nuclear Dynamics 4, ed. W. Bauer and H.-G.Ritter
Plenum Press.N.Y. page 267 (hep-ph/9806379).

\bibitem{kpp} B. Z. Kopeliovich, B.~Povh and E.~Predazzi,
Phys. Lett. {\bf B405} (1997) 361.

\bibitem{as} U.~Amaldi and K.R.~Schubert,
Nucl. Phys. {\bf B166} (1980) 301.

\bibitem{zkl} Al.B.~Zamolodchikov, B.Z.~Kopeliovich and
L.I.~Lapidus, Sov. Phys. JETP Lett. {\bf 33} (1981) 612.

\bibitem{soffer} C.~Bourrely et al., Phys. Rev. {\bf D26} (1982) 1781.

\bibitem{hir} B.Z.~Kopeliovich, Soft Component of Hard Reactions
and Nuclear Shadowing (DIS, Drell-Yan reaction, heavy quark production),
in proc. of the Workshop Hirschegg'95: Dynamical  
 Properties of Hadrons in Nuclear Matter, Hirschegg, January 16-21,1995,
ed. by H. Feldmeier and W. N\"orenberg, Darmstadt, 1995, p. 102
(hep-ph/9609385). 

\bibitem{knp} B.Z.~Kopeliovich, N.N. Nikolaev and I.K.~Potashnikova,
Phys. Lett. {\bf 209B} (1988) 335; Phys. Rev. {\bf D39} (1989) 769.

\bibitem{gribov} V.N.~Gribov, Eur. Phys. J. {\bf C10} (1999) 71; 
{\it ibid} {\bf C10}, 91 (1999).

\bibitem{ewerz} C.~Ewerz, Eur. Phys. J. {\bf C13} (2000) 503.

\bibitem{kmr} V.A.~Khoze, A.D.~Martin and M.G.~Ryskin, hep-ph/0007359

\bibitem{dmw} Yu.~Dokshitzer, G.~Marchesini, B.~Webber,
Nucl. Phys. {\bf B469}, 93 (1996).

\bibitem{gribov69} V.N.~Gribov,  Sov.  Phys.  JETP {\bf 57} (1969) 1306.    

\bibitem{dino} K.~Goulianos, J.~Montanha, Phys. Rev. {\bf D59} (1999)
114017.

\bibitem{schlein} S.~Erhan and P.E.~Schlein,
Phys. Lett. {\bf B427} (1998) 389.

\bibitem{kaidalov} A.B.~Kaidalov, Phys. Rept. {\bf 50}, 157 (1979). 

\bibitem{cdf} F.~Abe et al., Phys. Rev. {\bf D50}, 550 (1993). 

\bibitem{pdt1} C. Caso et al, The Eur. Phys. J. {\bf C3}, 1  (1998). 

\bibitem{bel} U.~Amaldi et al., Phys. Lett. {\bf 36B} (1971) 504;
{\bf 66B} (1977) 390;
M.~Ambrosio et al., Phys. Lett. {\bf 115B} (1982) 495;
N.~Amos et al., Phys. Lett. {\bf 128B} (1983) 343;
Nucl. Phys. {\bf B262} (1985) 689;
Phys. Rev. Lett. {\bf 61} (1988) 525;
Phys. Rev. Lett. {\bf 63} (1989) 2784;
V.~Apokin et al., Sov. J. Nucl. Phys. {\bf 25} (1977) 51;
V.~Bartenev et al., Phys. Rev. Lett. {\bf 29} (1972) 1755;
G.~Beznogikh et al., Nucl. Phys. {\bf B54} (1973) 78;
M.~Bozzo et al., Phys. Lett. {\bf 147B} (1984) 385;
A~.Breakstone et al., Nucl. Phys. {\bf B248} (1984) 253;
R.E.~Breedon et al., Phys. Lett. {\bf 216B} (1989) 459;
C.~Bromberg et al., Phys. Rev. {\bf D15} (1977) 64;
J.P.~Burq et al., Phys. Lett. {\bf 109B} (1982) 124;
R.L.~Cool et al., Phys. Rev. {\bf D24} (1981) 2821;
D.~Favart et al., Phys. Rev. Lett. {\bf 47} (1981) 1191.

\bibitem{data} R.~Battiston et al.,  Phys. Lett. {\bf 127B} (1983) 472;
M.~Bozzo et al., Phys. Lett. {\bf 147B} (1984) 385;  {\bf 155B} (1985) 197;
D.~Bernard et al., Phys. Lett. {\bf 198B} (1987) 583;
G.~Arnison et al., {\bf 128B} (1983) 336.

\bibitem{rho} U.~Amaldi et al., Nucl. Phys. {\bf 166B} (1980) 301;
C.~Augier et al., Phys. Lett. {\bf 316B} (1993) 448;
N.~Amos et al., Phys. Rev. Lett. {\bf 68} (1992) 2433.

\bibitem{lhd} A.K.~Likhoded and O.P.~Yushchenko,
Int. J. Mod. Phys. {\bf A6} (1991)913;
A.K. Likhoded, V.A. Uvarov, P.V. Chliapnikov, 
Phys. Lett. {\bf B215} (1988) 417

\bibitem{wang} X.-N.~Wang, Phys. Rept. {\bf 280} (1997) 287

\end{thebibliography}
\end{document}